\documentclass[nobibnotes,nofootinbib,prd]{revtex4}
\usepackage{amsmath,amssymb,bm}
\usepackage{epsfig}
\usepackage{graphicx}
\usepackage{pstricks,pst-node,pst-text,pst-3d}
\usepackage{subfigure}
\usepackage{hyperref}

\begin{document}
\title{ Inclusive gauge boson production in the color dipole formalism}

\author{E.\ A.\ F.\ Basso}
\email{eduardo.basso@thep.lu.se}
 \affiliation{Department of
Astronomy and Theoretical Physics, Lund University, SE-223 62 Lund,
Sweden}

 \author{V.\ P.\ Gon\c{c}alves}
 \email{barros@ufpel.edu.br}
 \affiliation{Instituto de F\'{\i}sica e Matem\'atica, Universidade Federal de Pelotas (UFPel),
Caixa Postal 354, CEP 96010-900, Pelotas, RS, Brazil.}

 \author{M. Rangel}
 \email{rangel@if.ufrj.br}
 \affiliation{Instituto de F\'{\i}sica, Universidade Federal do Rio de Janeiro (UFRJ), Rio de Janeiro, Brazil.}

\date{\today}

\begin{abstract}


The inclusive production of the gauge bosons $W^{\pm}$ and $Z^0$ is studied within the color dipole formalism. Gluon saturation effects associated to non-linear corrections to the QCD dynamics, which are expected to contribute at high energies, are naturally included in such formalism. We estimate the contribution of these effects at LHC energies and compare our results with the next-to-next-to-leading order collinear predictions. A comparison with the current experimental data is performed  and predictions for higher energies are presented.

\end{abstract}
\maketitle

\section{Introduction}
The production of the massive gauge bosons $W^{\pm}$ and $Z^0$ is one of the few process in $pp$ collisions where the collinear factorization has been rigorously proven and  the perturbative Quantum Chromodynamics (pQCD) predictions are known up to next-to-next-to-leading order (NNLO) \cite{nnlo}. Therefore, its study at Tevatron and LHC energies provide an important test of the Standard Model (SM)  as well as  on the pQCD and the higher order corrections.
In the collinear factorization approach the gauge boson production is viewed, at leading order (LO), as the fusion of the quark and antiquark which produces a gauge boson, being a unique process which offers high sensitivity to the parton distribution in the proton \cite{review_watt}. Recently, the ATLAS \cite{atlas_data}, CMS \cite{cms_data,cms_data_8tev} and LHCb \cite{lhcb_data} collaborations have performed $W$ and $Z$ precision measurement of the inclusive cross sections  at $\sqrt{s}$ = 7 and 8 TeV, with the data being well described by the NNLO pQCD predictions.
A major theoretical uncertainty on the cross sections predictions is due to uncertainties on the parton distribution functions (PDFs) \cite{watt2}. While its behaviour at high values of the Bjorken variable $x$  have been determined from fixed target and HERA data and confirmed at higher virtualities $Q^2$ by $W$ and $Z$ production at Tevatron, for smaller $x$ values, the PDFs have been measured by HERA alone but at much lower $Q^2$. For the energies probed at LHC, they must be evolved  using the DGLAP equations \cite{dglap}. In particular,  the measurement of the gauge boson cross section at the LHCb \cite{lhcb_data}, which probes forward rapidities,  have a sensitivity to values of $x$ as low as $1.7 \times 10^{-4}$ and values of  $Q^2 \approx M_G^2$, where $G = W^{\pm}$ or $Z^0$. Consequently, the current and future experimental data for the gauge boson production provide an important test of the QCD dynamics in a new kinematical range, where new dynamical effects can contribute.

One of the such effects  is the replacement of  the usual collinear approach \cite{collfact}  by a more general factorization scheme, as for example the   $k_{\perp}$-factorization  approach \cite{CCH,CE,GLRSS}. While in the collinear factorization approach \cite{collfact} all partons involved are assumed to be on mass shell, carrying only longitudinal momenta, and their transverse momenta are neglected in the QCD matrix elements, in the $k_{\perp}$-factorization  approach the effects of the finite transverse momenta of the incoming partons are taken into account. In this case 
the cross sections are now $k_{\perp}$-factorized into an off-shell partonic cross section and a  $k_{\perp}$-unintegrated parton density functions ${\cal{F}}_i(x,k_{\perp})$ \cite{CCH,CE,GLRSS}.    
A sizeable piece  of the NLO and some of the NNLO corrections to the LO contributions on the collinear approach, related to  the  contribution of non-zero transverse momenta of the incident partons, are already included in the LO contribution within the $k_{\perp}$-factorization approach. Moreover, the coefficient functions and the splitting functions giving the collinear parton distributions are supplemented by all-order $\alpha_s\ln (1/x)$ resummation at high energies \cite{CH}. 
Such approach was applied for the gauge boson production in the Refs. \cite{wmr,baranov,baranov2,deak,hautmann}.

Another possible dynamical effect is the gluon saturation associated to non-linear corrections to the QCD dynamics, which are expected to contribute at high energies (For recent reviews see Ref. \cite{hdqcd}). In
this regime, perturbative QCD  predicts that
the small-$x$ gluons in a hadron wavefunction should form a Color
Glass Condensate (CGC) \cite{CGC}, which is characterized by the limitation on the maximum
phase-space 
allowed in the hadron
wavefunction (parton saturation). The transition is then
specified  by a typical energy dependent scale, called saturation scale $Q_{\mathrm{sat}}$. 
One of the main implications of the gluon saturation is that it leads to the breakdown of the twist expansion and the factorization schemes \cite{raju}.

\begin{figure}[t]
\centering
\subfigure[]{
\scalebox{0.3}{\includegraphics{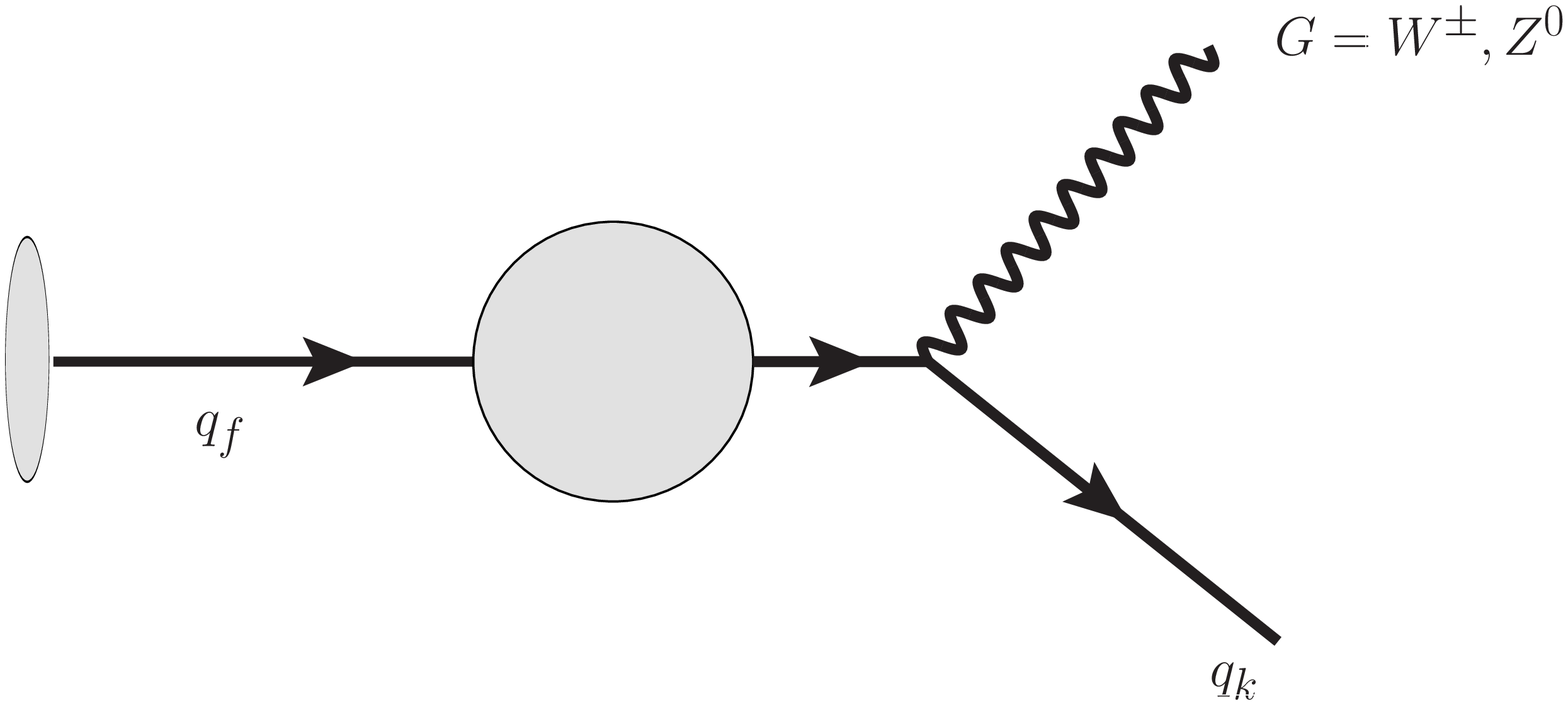}}
\label{fig:gb_dir}
}
\centering
\subfigure[]{
\scalebox{0.3}{\includegraphics{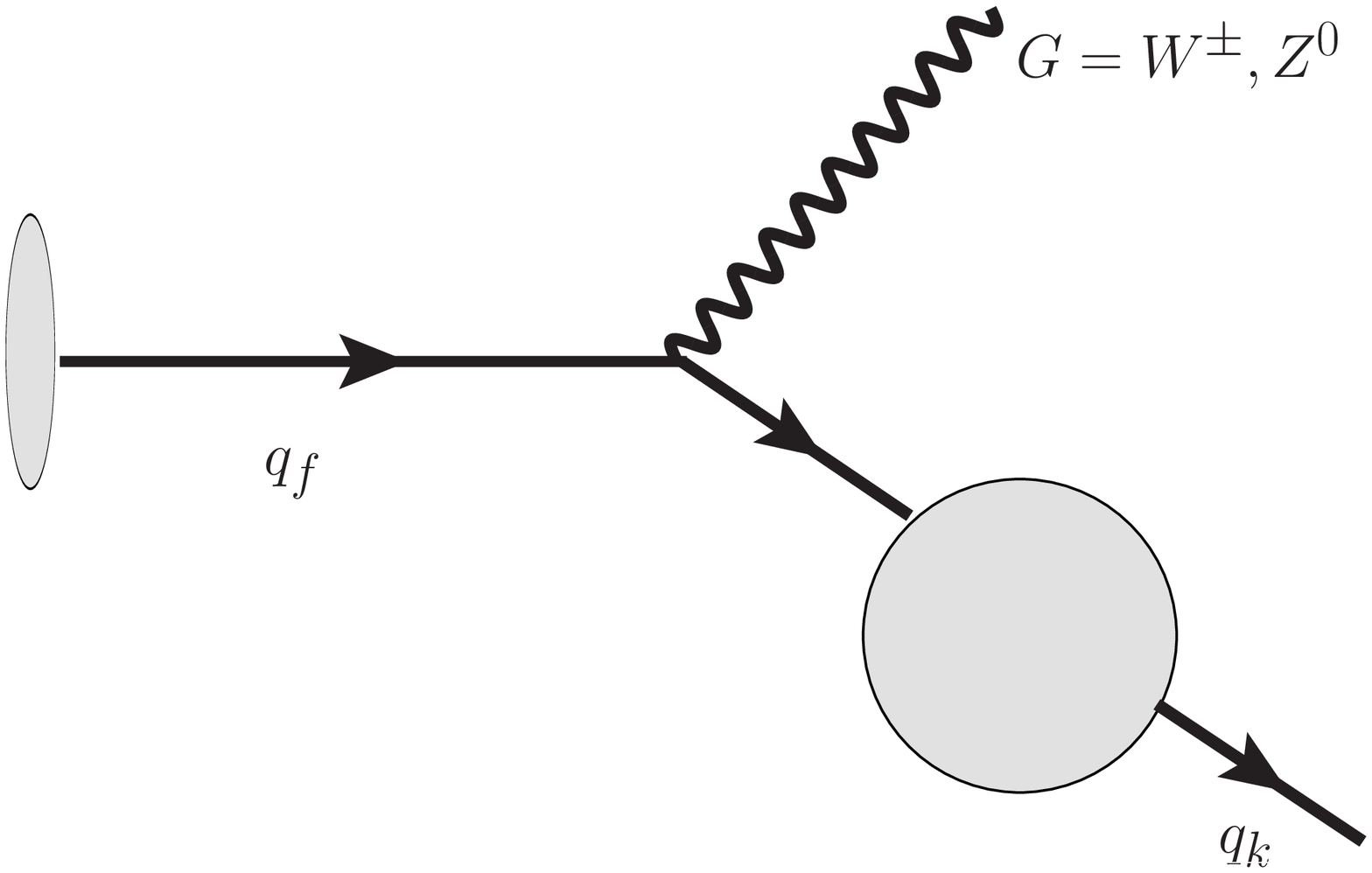}}
\label{fig:gb_frag}
}
\caption{Diagrams contributing to the process of a gauge boson irradiated by a quark (antiquark) of flavor $f$ either (a) after  or (b) before  the interaction with the target color field (denoted by a shaded circle). For the $W^{\pm}$ radiation one have $q_k \neq q_f$. }
\label{fig:gb_dip}
\end{figure}

Gluon saturation effects can be naturally described in the color dipole formalism \cite{nik}. At high energies color dipoles with a defined 
transverse separation are  eigenstates of the interaction. The main quantity in this formalism is the dipole-target cross section, which is universal and determined by  QCD dynamics at high energies. In particular, it provides a unified description of inclusive and diffractive observables in $ep$ processes as well as for in  Drell-Yan, prompt photon and heavy quark production in hadron-hadron collisions \cite{nik,nik_dif,k95,kts99,brodsky,krt01,npz}. The basic idea present in the description of hadronic collisions using the  color dipole formalism  is that although cross sections are Lorentz invariant, the partonic interpretation of the microscopic process depends on the reference frame \cite{k95}. In particular, in the target rest frame,  the gauge boson production  can be described in the dipole formalism
as a bremsstrahlung process, rather than parton annihilation, with the  space-time picture being illustrated in Fig. \ref{fig:gb_dip}. A quark (or an antiquark) of the projectile hadron radiates a gauge boson. The radiation can occur after or before the quark scatters off the target. 
When the energy is high, the projectile quark is probing potentially dense gluon fields in the target, which implies that multiple scatterings have to be taken into account.
In this paper we consider the extension of  this formalism for the inclusive massive gauge boson production, as derived in Ref. \cite{pkp}, and calculate the cross sections considering the 
impact parameter dependent color glass condensate (bCGC) dipole model proposed in Ref. \cite{kmw}, which is based on the Balitsky - Kovchegov non-linear evolution equation \cite{BAL,kov} and takes into account the impact-parameter dependence of the saturation scale. As demonstrated in Ref. \cite{amir}, this model is able to describe  the recently released high precision combined HERA data. We estimate the magnitude of the non-linear effects at LHC and compare our predictions with those obtained using the collinear factorization and DGLAP equation. Moreover, our predictions are compared with the recent experimental data.

This paper is organized as follows. In the next Section, we present a brief review of the description of gauge boson production in the color dipole formalism. In Section \ref{dinamica} we present the main aspects of the QCD dynamics at high energies and the models for the scattering amplitude used in our calculations. In Section \ref{resultados} we present our predictions for the total cross sections. In particular, we estimate the magnitude of the non-linear effects at LHC energies. A comparison with the collinear predictions and experimental data is performed. Finally, in Section \ref{conclusoes}, our main conclusions are summarized.

\section{Gauge boson production in the color dipole picture}
\label{formalismo}

In the color dipole picture, the production mechanism of a gauge boson looks like a bremsstrahlung \cite{k95,kts99,brodsky} as represented in Fig. \ref{fig:gb_dip}. In the high energy limit, each of the two graphs factorizes into a production vertex for the gauge boson times an amplitude for scattering a quark off the target. The quark scatters at different impact parameters depending on whether the gauge boson is irradiated after of before the scattering. The interference between these scattering amplitudes implies that  the squared matrix element for  the gauge boson production is expressed  in terms of the same dipole - target cross section  $ \sigma_{q\bar{q}}$ determined by the low-$x$ DIS data. In particular, the cross section for the radiation of a gauge boson $G$ of mass $M$ and transverse ($T$) or longitudinal ($L$) polarization from a fast quark of flavor $f$ and mass $m_f$, which takes a fraction $\alpha$ of the radiating quark energy is given by
\begin{eqnarray}
\sigma^f_{T,L}(qN \rightarrow G X) = \int \frac{d \alpha}{\alpha} \int d^2 \rho |\Psi^{V - A}_{T,L}(\alpha,\rho,M,m_f)|^2 \sigma_{q\bar{q}}(\alpha \rho,x)
\end{eqnarray}
where $\rho$ is the transverse separation between $G$ and the final quark and the light-cone wave function $\Psi$ describes the electroweak radiation $q \rightarrow G q$, which can be a vector or axial-vector transition. Moreover, $\sigma_{q\bar{q}}$ is the dipole - target cross section, which is determined by the QCD dynamics at high energies to be discussed in the next section, with $x$ being the Bjorken variable which is directly related to the energy scale.
The above equation can be generalized in order to estimate the transverse momentum distribution of the gauge bosons, being given by
\begin{eqnarray}
\frac{d\sigma^f_{T,L} (qN \rightarrow GX)}{d \ln \alpha d^2q_{\perp}} & = & \frac{1}{(2\pi)^2} \int d^2\rho_1 d^2\rho_2 \exp[i\vec{q}_{\perp} \cdot (\vec{\rho}_1 - \vec{\rho}_2)] \Psi^{V - A}_{T,L}(\alpha,\vec{\rho}_1,M,m_f) \Psi^{V - A, *}_{T,L}(\alpha,\vec{\rho}_2,M,m_f) \nonumber \\
 & \times & \frac{1}{2}\left[ \sigma_{q\bar{q}}(\alpha \vec{\rho}_1,x) + \sigma_{q\bar{q}}(\alpha \vec{\rho}_2,x ) - \sigma_{q\bar{q}}(\alpha|\vec{\rho}_1- \vec{\rho}_2|,x)\right] \,\,,
\end{eqnarray}
where $q_{\perp}$ is the transverse momentum of the outgoing gauge boson and $\vec{\rho}_1$ and $\vec{\rho}_2$ are the quark - gauge boson transverse separations in the two radiation amplitudes contributing to the cross section. 
  For an unpolarized initial quark the vector and axial-vector contributions to the wavefunctions are decorrelated, {\it i.e.}, their interference is destructive, in a way that \cite{pkp}
\begin{eqnarray}
&  &\sum_\text{quark pol.} \Psi^{V-A}_{T,L}(\alpha,\vec{\rho}_1,M,m_f) \Psi^{V-A}_{T,L}(\alpha,\vec{\rho}_2,M,m_f)  = \nonumber \\ 
& = & \Psi^V_{T,L}(\alpha,\vec{\rho}_1,M,m_f)\Psi^V_{T,L}(\alpha,\vec{\rho}_2,M,m_f) + \Psi^A_{T,L}(\alpha,\vec{\rho}_1,M,m_f) \Psi^A_{T,L}(\alpha,\vec{\rho}_2,M,m_f)\,\,.
\end{eqnarray}
The different $\Psi$ components are written as 
\begin{eqnarray}\label{VV}
&&\Psi^{T}_{V}(\alpha,\vec{\rho}_1)
\Psi^{T*}_{V}(\alpha,\vec{\rho}_2)= \frac{{\cal C}_f^2(g^{G}_{v,f})^2}{2\pi^2}\Bigg\{
     m_f^2 \alpha^4 {\rm K}_0\left(\eta \rho_1\right)
     {\rm K}_0\left(\eta \rho_2\right)+ \left[1+\left(1-\alpha\right)^2\right]\eta^2
   \frac{\vec{\rho}_1\cdot\vec{\rho}_2}{\rho_1 \rho_2}
     {\rm K}_1\left(\eta \rho_1\right)
     {\rm K}_1\left(\eta \rho_2\right)\Bigg\},\nonumber \\ 
&&\Psi^{L}_{V}(\alpha,\vec{\rho}_1)
\Psi^{L*}_{V}(\alpha,\vec{\rho}_2)=
\frac{{\cal C}_f^2(g_{v,f}^G)^2}{\pi^2}M^2
\left(1-\alpha\right)^2
  {\rm K}_0\left(\eta \rho_1\right)
     {\rm K}_0\left(\eta \rho_2\right)\,, \nonumber \\
&&\Psi^{T}_{A}(\alpha,\vec{\rho}_1)\Psi^{T*}_{A}(\alpha,\vec{\rho}_2)=
   \frac{{\cal
   C}_f^2(g_{a,f}^G)^2}{2\pi^2}\Bigg\{
     m_f^2 \alpha^2(2-\alpha)^2 {\rm K}_0\left(\eta \rho_1\right)
     {\rm K}_0\left(\eta \rho_2\right)+ \left[1+\left(1-\alpha\right)^2\right]\eta^2
   \frac{\vec{\rho}_1\cdot\vec{\rho}_2}{\rho_1 \rho_2}
     {\rm K}_1\left(\eta \rho_1\right)
     {\rm K}_1\left(\eta \rho_2\right)\Bigg\}, \nonumber \\
&&\Psi^{L}_{A}(\alpha,\vec{\rho}_1)
\Psi^{L*}_{A}(\alpha,\vec{\rho}_2) = \frac{{\cal
C}_f^2(g_{a,f}^G)^2}{\pi^2}\frac{\eta^2}{M^2}\Bigg\{\eta^2
  {\rm K}_0\left(\eta \rho_1\right)
     {\rm K}_0\left(\eta \rho_2\right)+\alpha^2m_f^2\frac{\vec{\rho}_1\cdot\vec{\rho}_2}{\rho_1 \rho_2}
     {\rm K}_1\left(\eta \rho_1\right)
     {\rm K}_1\left(\eta \rho_2\right)\Bigg\}, \nonumber
\end{eqnarray}
where $\eta = (1-\alpha)M^2 + \alpha^2 m_f^2$ and the coupling factors ${\cal C}_f$ are 
\begin{eqnarray} \nonumber
 {\cal
C}^Z_f=\frac{\sqrt{\alpha_{em}}}{\sin 2\theta_W},\qquad {\cal
C}^{W^+}_f=\frac{\sqrt{\alpha_{em}}}{2\sqrt{2}\sin\theta_W}V_{f_uf_d},\qquad
{\cal
C}^{W^-}_f=\frac{\sqrt{\alpha_{em}}}{2\sqrt{2}\sin\theta_W}V_{f_df_u}\,,
\end{eqnarray}
with the vectorial coupling at LO being given by
\begin{eqnarray}\label{vec}
g_{v,f_u}^Z=\frac12-\frac43\sin^2\theta_W,\qquad
g_{v,f_d}^Z=-\frac12+\frac23\sin^2\theta_W,\qquad g_{v,f}^W=1\,,
\end{eqnarray}
and 
\begin{eqnarray}\label{axial}
g_{a,f_u}^Z=\frac12,\qquad g_{a,f_d}^Z=-\frac12\,,\qquad
g_{a,f}^W=1\,
\end{eqnarray}
in the axial-vector case.

In order to estimate the hadronic cross section for the inclusive process $p p \rightarrow G X$ one has to note that the gauge boson carries away the momentum fraction $x_1$ from the projectile proton, with the light-cone fraction of the quark emitting the gauge boson $x_q$ being given by $x_q = x_1/\alpha$. Taking into account that the probability to find a quark (antiquark) with momentum fraction $x_q$ in the proton wave function is described in terms of the quark densities $q_f$ ($\bar{q}_f$) results that the cross section for the inclusive  production of a gauge boson of mass $M$ and transverse momentum $q_{\perp}$ is expressed as follows
 \cite{pkp}
 \begin{eqnarray}\label{eq:gb_qt}
\frac{d^4 \sigma_{T,L} (pp \rightarrow GX)}{d^2q_\perp dx_1}  &= & \frac{1}{2\pi}\sum_f \int_{x_1}^1 \frac{d\alpha}{\alpha^2} \left[ q_f(x_q,\mu_F^2) + {\bar{q}_f}(x_q,\mu_F^2) \right] \int d^2\rho_1 d^2\rho_2 \exp[i \vec{q}_\perp\cdot(\vec{\rho_1}-\vec{\rho_2})] \nonumber \\
&\times& \Psi_{T,L}^{V-A}(\vec{\rho_1}, \alpha) \Psi_{T,L}^{V-A}(\vec{\rho_2}, \alpha) \frac{1}{2}[\sigma_{q\bar{q}}(\alpha \rho_1,x) + \sigma_{q\bar{q}}(\alpha \rho_2,x) - \sigma_{q\bar{q}}(\alpha|\vec{\rho_1} - \vec{\rho_2}|,x)] \,\,,
\end{eqnarray}
with the Bjorken variable $x$ being given by $x = M^2/\hat{s}$, where $\hat{s} = s x_1/\alpha$ is the quark - proton center-of-mass energy squared and $\mu_F$ is the factorization scale.

It is important to emphasize that although both valence and sea quarks in the projectile are taken into account through the parton distributions, the color dipole accounts only for pomeron exchange from the target, disregarding its valence content. Therefore, in principle this approach is well suited for high energies and, consequently, small values of $x$. 

 \section{QCD Dynamics}
 \label{dinamica} 

In the color dipole formalism the   cross sections are determined by the dipole - proton cross section, $\sigma_{q\bar{q}}$, which  encodes all the information about the hadronic scattering, and thus about the non-linear and quantum effects in the hadron wave function. It can be expressed by
\begin{equation}\label{eq:dipcross}
\sigma_{q\bar{q}}(\vec{\rho},x)=2\int d^2b\, {\cal{N}} (\vec{b},\vec{\rho},Y),
\end{equation}
where ${\cal{N}}(\vec{b},\vec{\rho},Y)$ is the imaginary part of the forward amplitude for the scattering between a small dipole
(a colorless quark-antiquark pair) and a dense hadron target, at a given
rapidity interval $Y=\ln(1/x)$. The dipole has transverse size given by the vector
$\vec{\rho}=\vec{x}-\vec{y}$, where $\vec{x}$ and $\vec{x}$ are the transverse vectors for the quark
and antiquark, respectively, and impact parameter $\vec{b}=(\vec{x}+\vec{y})/2$. 
At high energies the evolution with the rapidity $Y$ of
${\cal{N}}(\vec{b},\vec{\rho},x)$  is given by the infinite hierarchy of equations, the so called
Balitsky-JIMWLK equations \cite{BAL,CGC}, which reduces in the mean field approximation to the Balitsky-Kovchegov (BK) equation \cite{BAL,kov}. 
In recent years,  the running coupling corrections to BK evolution kernel was explicitly calculated  \cite{kovwei1,balnlo},  including the  $\alpha_sN_f$ corrections to 
the kernel to all orders, and its solution studied in detail  \cite{javier_kov,javier_prl}. Basically, one has that the running of the coupling reduces the speed of the evolution to values compatible with experimental $ep$ HERA data \cite{bkrunning,weigert}.  The numerical solutions of the running coupling BK equation presented in Refs. \cite{bkrunning,weigert} assumed the translational invariance approximation, which implies ${\cal{N}}(\vec{b},\vec{\rho},Y) = {\cal{N}}(\vec{\rho},Y) S(\vec{b})$, with the normalization of the dipole cross section being fitted to data.
Unfortunately, impact-parameter dependent numerical solutions to the BK equation are very difficult to obtain \cite{stasto}. Moreover, the choice of the impact-parameter profile of the dipole amplitude entails intrinsically nonperturbative physics, which is beyond the QCD weak coupling approach of the BK equation. In fact, the BK equation generates a power law Coulomb-like tail, which is not confining at large distances and therefore can violate the unitarity bound. 
 It is important to emphasize that although a complete analytical solution of the BK equation is still lacking, its main properties at fixed $\vec{b}$ are known: (a) for the interaction of a small dipole ($\rho \ll
1/Q_{\mathrm{sat}}$), ${\cal{N}}(\vec{b},\vec{\rho},Y) \approx \rho^2$, implying  that
this system is weakly interacting; (b) for a large dipole ($\rho \gg
1/Q_{\mathrm{sat}}$), the system is strongly absorbed and therefore
${\cal{N}}(\vec{b},\vec{\rho},Y) \approx 1$. The typical momentum scale, $Q_{\mathrm{sat}}^2\propto x^{-\lambda}\,(\lambda\approx 0.3)$, is the so called saturation scale. This property is associated  to the
large density of saturated gluons in the hadron wave function. In the last years, several groups have constructed phenomenological models which satisfy the asymptotic behaviours of the BK equation in order to fit the HERA and RHIC data (See e.g. Refs. \cite{GBW,iim,kkt,dhj,Goncalves:2006yt,buw,kmw,agbs}).
In particular, in Ref. \cite{kmw} the authors have proposed  a generalization of the Color Class Condensate (CGC) model \cite{iim} with  the inclusion of   the impact parameter dependence in the dipole - proton scattering amplitude.
  This impact parameter dependent color glass condensate (bCGC) dipole model is based on the Balitsky - Kovchegov non-linear evolution equation and improves the CGC model by incorporating the impact-parameter dependence of the saturation scale.
The corresponding dipole - proton scattering is given by 
\cite{kmw} 
\begin{eqnarray}
\mathcal{N}(\vec{b},\vec{\rho},Y) =   
\left\{ \begin{array}{ll} 
{\mathcal N}_0\, \left(\frac{ \rho \, Q_{\mathrm{sat}}}{2}\right)^{2\left(\gamma_s + 
\frac{\ln (2/\rho Q_{\mathrm{sat}})}{\kappa \,\lambda \,Y}\right)}  & \mbox{$\rho Q_{\mathrm{sat}} \le 2$} \\
 1 - \exp^{-A\,\ln^2\,(B \, \rho \, Q_{s,p})}   & \mbox{$\rho Q_{\mathrm{sat}}  > 2$} 
\end{array} \right.
\label{eq:bcgc}
\end{eqnarray}
with  $Y=\ln(1/x)$ and $\kappa = \chi''(\gamma_s)/\chi'(\gamma_s)$, where $\chi$ is the 
LO BFKL characteristic function.  The coefficients $A$ and $B$  
are determined uniquely from the condition that $\mathcal{N}(\vec{b},\vec{\rho},Y)$, and its derivative 
with respect to $\rho Q_{\mathrm{sat}}$, are continuous at $ \rho Q_{\mathrm{sat}}=2$. 
In this model, the proton saturation scale $Q_{\mathrm{sat}}$ depends on the impact parameter:
\begin{equation} 
  Q_{\mathrm{sat}}\equiv Q_{\mathrm{sat}}(x,\vec{b})=\left(\frac{x_0}{x}\right)^{\frac{\lambda}{2}}\;
\left[\exp\left(-\frac{{b}^2}{2B_{\rm CGC}}\right)\right]^{\frac{1}{2\gamma_s}}.
\label{newqs}
\end{equation}
The parameter $B_{\rm CGC}$  was  adjusted to give a good 
description of the $t$-dependence of exclusive $J/\psi$ photoproduction.  
Moreover the factors $\mathcal{N}_0$ and  $\gamma_s$  were  taken  to be free. In this way a very good description of  $F_2$ data was obtained. 
Recently this model has been improved by the fitting of its free parameters considering the high precision combined HERA data \cite{amir}. The set of parameters  which will be used here are the following: $\gamma_s = 0.6599$, $B_{CGC} = 5.5$ GeV$^{-2}$, $\mathcal{N}_0 = 0.3358$, $x_0 = 0.00105 \times 10^{-5}$ and $\lambda = 0.2063$. 
The value of $\gamma_s$ deserves some comments. In the CGC model, in which the impact parameter dependence of the saturation scale is disregarded,  the authors have assumed a fixed value of the anomalous dimension, $\gamma_s = 0.63$, which is the value predicted by the BFKL dynamics at leading order. Moreover, the old HERA data have been fitted considering only three light flavours. After, in Ref. \cite{Soyez2007}, the charm contribution was included in the analysis assuming $\gamma_s$ as a free parameter. In this case, a value of $\gamma_s = 0.7376$ was obtained. As demonstrated in \cite{amir}, a slightly higher value for $\gamma_s = 0.76$ is favoured by the recent HERA data.
Such higher values of $\gamma_s$ are rather close to what we expect from NLO BFKL, which implies $\gamma_s \ge 0.7$ \cite{Soyez2007}. In contrast, as demonstrated in Refs. \cite{kmw,amir}, the inclusion of the impact parameter dependence in the saturation scale reduces the extracted value of the anomalous dimension. The main distinction between these two studies is that the analysis performed in Ref. \cite{amir} using the recent HERA data seem to favour  a larger value than that obtained in Ref. \cite{kmw} using the old HERA data, with the value obtained in \cite{amir} being close to the LO BFKL value. Another important difference between these previous models and the bCGC model used in our studies is related to the distinct prediction for the energy behaviour of the saturation scale. As demonstrated in Ref. \cite{amir}, the recent HERA data implies a steeper energy dependence in comparison to that obtained considering the old HERA data, which is directly associated to the larger value for the parameter $\lambda = 0.206$ in comparison to the previous one ($\lambda = 0.119$). At central collisions $b=0$, the predictions of the CGC and rcBK dipole models are steeper than the bCGC one (See Figs. 2 and 3 in Ref. \cite{amir}), with $Q_s^2$ being predicted by the bCGC model to be $\approx$ 2 GeV$^2$ at $x \lesssim  10^{-6}$.





For comparison, in what follows we also will use the GBW model \cite{GBW}, which  assume ${\cal{N}}(\vec{b},\vec{\rho},Y) = {\cal{N}}(\vec{\rho},Y) S(\vec{b})$
with the scattering amplitude  parametrized
as follows
\begin{equation}\label{eq:gbw}
{\cal{N}}_{GBW}(\rho,Y)=1-e^{-\rho^2Q_{\mathrm{sat}}^2(Y)/4},
\end{equation}
where the saturation scale is given by $Q_{\mathrm{sat}}^2=Q_0^2\left(x_0/x\right)^{\lambda}$,
$x_0$ is the value of the Bjorken $x$ in the beginning of the evolution and $\lambda$ is the
saturation exponent. In comparison to the bCGC model, the GBW model implies a steeper energy dependence for the saturation scale, with $Q_s^2$ being  $\approx$ 4 GeV$^2$ at $x \lesssim  10^{-6}$.
Our motivation to use this phenomenological model is associated to the fact that its linear limit is known, being given by 
\begin{eqnarray}
{\cal{N}}_{GBW - lin}(\rho,Y)=\frac{\rho^2Q_{\mathrm{sat}}^2(Y)}{4} \,\,.
\end{eqnarray}
It allows to compare the full and linear predictions for the gauge boson production cross sections and thus estimate the magnitude of the non-linear corrections for these observables.

\begin{figure}[t]
\large
\begin{center}
\scalebox{0.9}{\includegraphics{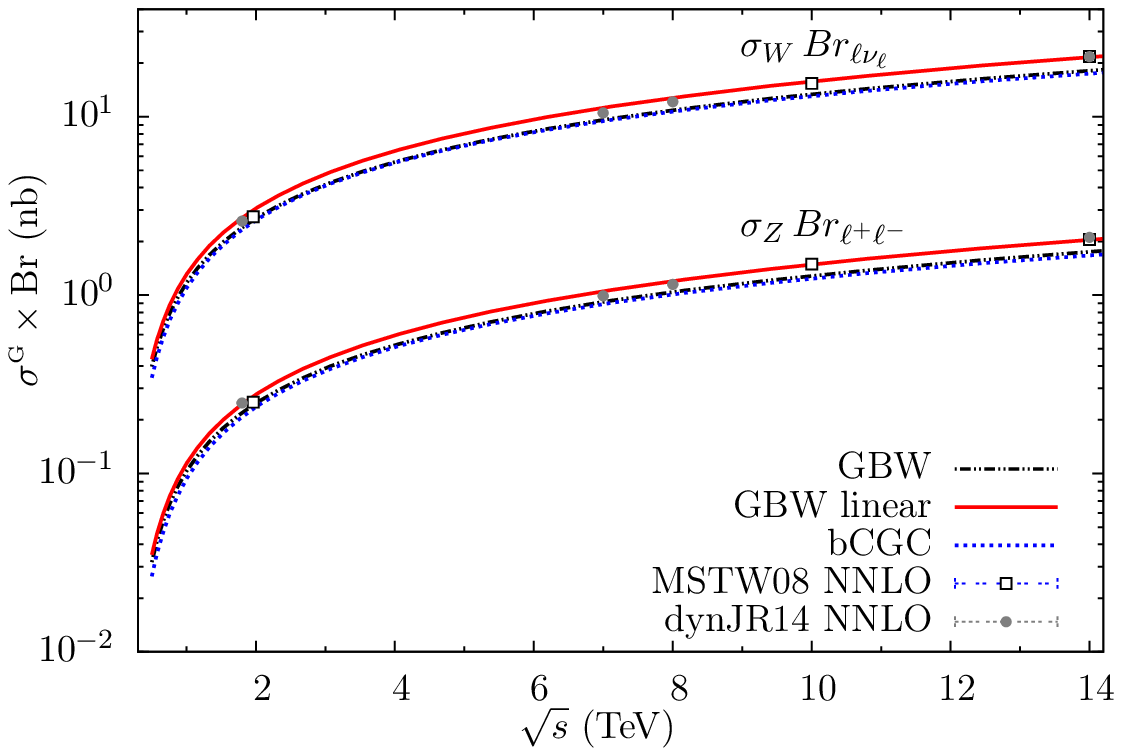}}
\caption{(Color online) Energy dependence of the total cross sections predicted by the  linear (GBW Linear)  and non-linear (GBW and CGC) models. The NNLO collinear predictions obtained in Refs. \cite{mstw,jr} are presented for comparison.  }
\label{fig:1}
\end{center}
\end{figure}
\normalsize

\section{Results}
\label{resultados}
In what follows we present our predictions for the $W$ and $Z$-boson production obtained using the color dipole formalism and the three phenomenological models for the  dipole - proton scattering amplitude discussed above. Following Ref. \cite{GBW} we will assume that the quark masses are given by $m_u = m_d = m_s = 0.14$ GeV, $m_c = 1.4$ GeV and $m_b = 4.5$ GeV. Moreover, we will assume that the factorization scale is equal to the gauge boson mass and use the CTEQ6L  parameterization \cite{cteq} for the parton distribution functions. We have verified that our results do not depend on these choices. It is important to emphasize that the expression for the gauge boson production in the color dipole formalism was derived at leading order in $\log (1/x)$, disregarding the contributions of valence quarks and larger-$x$ corrections. Such contributions can modify the normalization and the behaviour at low energies of our predictions.  A possibility to estimate the magnitude of these corrections is to leave the normalization of cross sections as being a free parameter to be fixed by the experimental data, usually denoted $K$-factor. However, as in previous calculations for the Drell-Yan \cite{rauf} and heavy quark \cite{nos} production  using the color dipole formalism
we will not include a $K$-factor. Consequently, our predictions will be parameter free. For comparison, we will compare our results with the NNLO collinear predictions available in the literature \cite{mstw,jr,mcfm} and with the current experimental data \cite{ua1,ua2,cdf1,cdf2,d0,cms_data,atlas_data,cms_data_8tev,lhcb_data}. 
In particular, we compare our predictions with the recent data from the ATLAS and CMS Collaborations, which have performed  $W$ and $Z$-boson measurements in the inclusive Drell-Yan processes  to high precision  for $pp$ collisions at 7 and 8 TeV. These  measurements have been made for the decay channels $W \rightarrow l \nu$ and $Z \rightarrow l\bar{l}$, with $l$ denoting a lepton, for the available phase volume  and extrapolated to the entire kinematical region.

\begin{table}[t]
\footnotesize 
\begin{center}
\begin{tabular}{|c@{\quad}||c@{\quad}|c@{\quad}|c@{\quad}|c@{\quad}|c@{\quad}|c@{\quad}|}
\hline

 \bf{$\sqrt{s}$ (TeV)} & {\bf GBW}  & {\bf GBW linear}  & {\bf bCGC}  & {\bf MSTW} & {\bf dynJR14} & {\bf DATA  (nb)}  \\ [0.5ex] \hline \hline

 \multicolumn{7}{|c|}{{\bf $Z_0 \rightarrow \ell^+ \, \ell^-$}} \\ \hline \hline

     1.8         & 0.21  &  0.25  &  0.21  & -- & $0.2456$ & \parbox[t]{5cm}{$0.231 \pm 0.012$ (CDF) \\ $0.221 \pm 0.012$ (D0) }\\ \hline      


     1.96        & 0.23  &  0.28  &  0.23  & $ 0.2507 $  &  --  & $0.2549 \pm 0.016$ (CDF) \\ \hline

     7           & 0.89  &  1.08     &  0.89  & -- & $0.9689$  & \parbox[t]{5cm}{$0.937 \pm 0.037$ (ATLAS) \\ $0.974 \pm 0.044$ (CMS) } \\ \hline
  
 
     8           & 1.02  &  1.23   &  1.00   & -- & $1.1271$  & $1.15 \pm 0.37$ (CMS) \\ \hline

     10          & 1.26  &  1.47   &  1.23   & $1.429$  &  --  & -- \\ \hline
     
     14          & 1.73  &  1.91   &   1.65  & $2.051$  & $2.0658$ & -- \\ \hline \hline

 \multicolumn{7}{|c|}{$W^+ + W^- \rightarrow \ell \, \mu_\ell$} \\ \hline \hline

     1.8         & 2.40  &  2.80  &  2.34  & -- & $2.5659$ & \parbox[t]{5cm}{$2.49 \pm 0.12$ (CDF) \\ $2.310 \pm 0.11$ (D0) }\\ \hline 


     1.96        & 2.64   &  3.08   &  2.59   & $ 2.747$ &  -- & $ 2.749 \pm 0.17$ (CDF)\\ \hline

     7           & 9.62   &  11.56 &  9.46   & -- & $10.2976$ & \parbox[t]{5cm}{$10.207	 \pm 0.403$ (ATLAS)\\ $10.3 \pm 0.43$ (CMS)}\\ \hline
 
 
     8           & 10.91 &  13.14  &  10.70 & -- & $11.8966$ & $12.21 \pm 0.40$ (CMS) \\ \hline

     10          & 13.39  &  16.24  &  13.07 & $ 15.35  $ &  --  & -- \\ \hline

     14          & 18.16 &  22.21   &  17.51 & $ 21.72$  & $21.32$ & -- \\ \hline

\end{tabular}
\end{center}
\caption{Comparison between the linear and non-linear predictions for the total cross sections for different values of the center-of-mass energy. The NLLO predictions obtained in Refs. \cite{mstw,jr} are also presented as well as the experimental results obtained in Refs.\cite{cdf1,cdf2,d0,cms_data,atlas_data, cms_data_8tev}.  Cross sections in nb. }
\label{tab-1}
\end{table}

\begin{figure}[t]
\large
\begin{center}
\scalebox{0.9}{\includegraphics{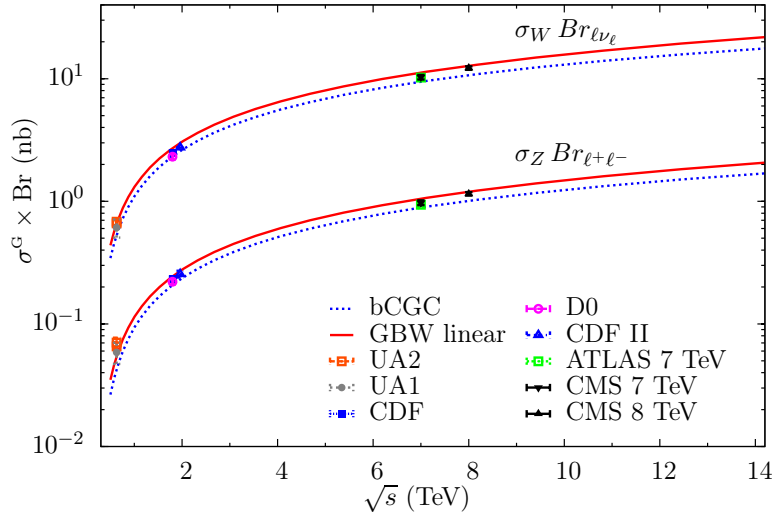}}
\caption{ (Color online) Energy dependence of the total cross section predicted by the bCGC and GBW linear models. Data from \cite{ua1,ua2,cdf1,cdf2,d0,cms_data,atlas_data}.}
\label{fig:2}
\end{center}
\end{figure}

In Fig. \ref{fig:1} we present our predictions for the energy dependence of the total cross sections. We observe that the GBW and bCGC predictions are similar. The GBW Linear predictions, which disregard gluon saturation effects, are  $\approx 20 \%$ larger than  the non-linear predictions.  In comparison with the NNLO collinear predictions presented in Refs. \cite{mstw,jr}, we obtain that the GBW Linear model predict similar values for the total cross sections. Our results indicate that the NNLO predictions  can be quite well  reproduced   using the color dipole formalism with $K = 1$ if the non-linear effects are disregarded, {\it i.e.} if we assume that the behaviour $\rho^2 x^{-\lambda}$ for the dipole - proton scattering amplitude is valid for all values of $\rho$. 
  A more detailed comparison can be made analysing the Table \ref{tab-1}, where we present the results for some typical values of the center-of-mass energy.  Moreover, in the Fig. \ref{fig:2} we present  the energy dependence of our predictions and provide   a comparison with  the current experimental data \cite{ua1,ua2,cdf1,cdf2,d0,cms_data,atlas_data,cms_data_8tev}(See also Table \ref{tab-1}).  We observe that the GBW and bCGC predictions are below of the experimental data. Considering that the bCGC describe quite well the HERA data, we can interpreted this result as an indication that the DGLAP evolution, disregarded in this model, is important for the description of the gauge boson production or that higher order corrections and/or valence quarks contributions for the color dipole formalism should be included in our calculations.

\begin{figure}[t]
\large
\begin{center}
\scalebox{0.9}{\includegraphics{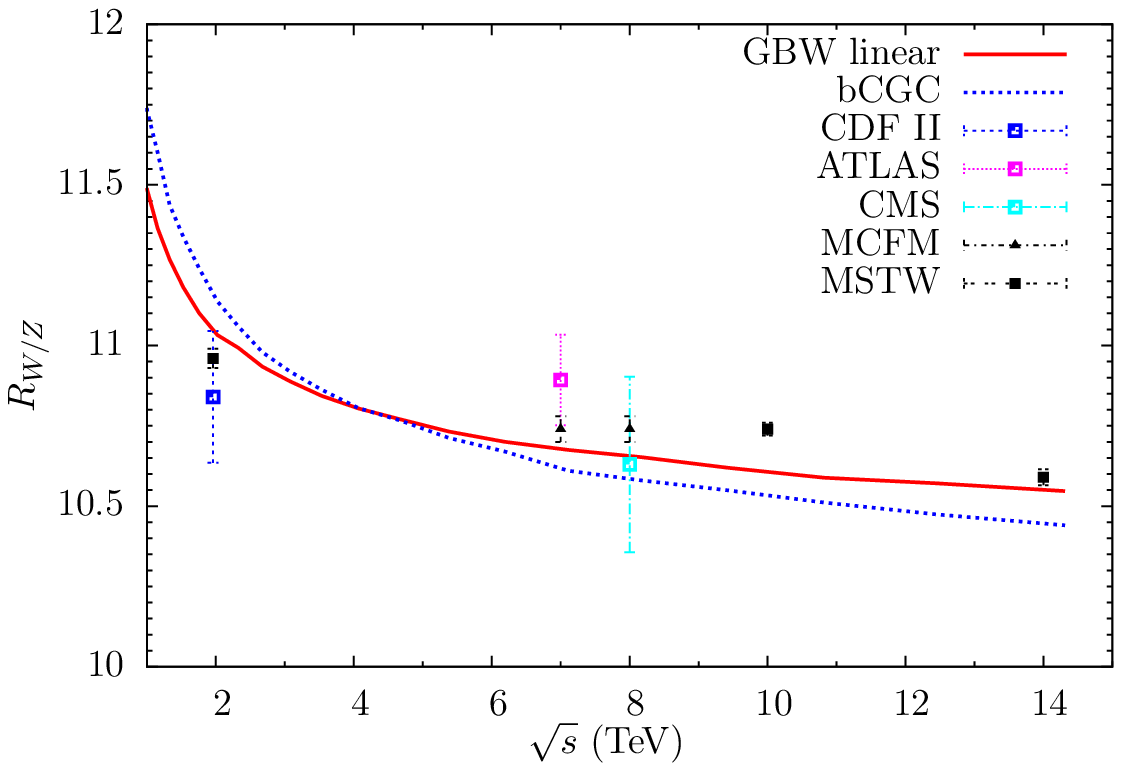}}
\caption{ (Color online) Energy dependence of the ratio between the $W$ and $Z$ cross sections. Data from \cite{cdf2,cms_data,atlas_data, cms_data_8tev}. The MCFM and MSTW NNLO predictions are presented for comparison.}
\label{fig:3}
\end{center}
\end{figure}

In Fig. \ref{fig:3} we compare our predictions for the ratio between the $W$ and $Z$ cross sections, denoted $R_{W/Z}$, with the experimental data from the CDF, ATLAS and CMS  Collaborations \cite{cdf2,atlas_data,cms_data_8tev} and theoretical NNLO predictions \cite{mstw,mcfm}. We obtain that our predictions strongly increases  at small values of energy, in disagreement with the data and the theoretical expectations for $R_{W/Z}$ in this kinematical regime. This behaviour is directly associated to the limitation of the color dipole formalism, which disregarded the valence quark contribution in the gauge boson cross section. In contrast, at larger values of energy, the GBW Linear prediction is similar to the NNLO one, with the bCGC one being smaller by $\approx 18 \%$, in agreement with our previous results for the total cross sections.

\normalsize

\normalsize

\begin{figure}[t]
\large
\begin{center}
\scalebox{0.9}{\includegraphics{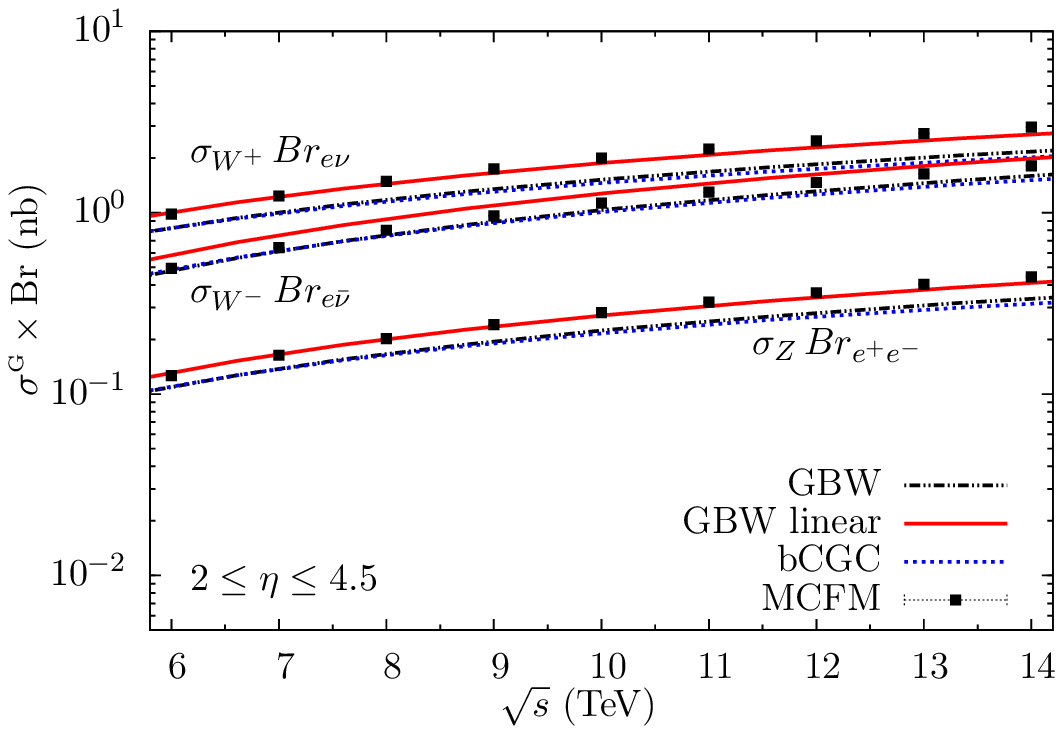}}
\caption{ (Color online) Energy dependence of the total cross sections obtained assuming that the gauge boson $G$ is produced in  the LHCb kinematical range ($ 2 \le \eta (G) \le 4.5$).}
\label{fig:4}
\end{center}
\end{figure}

\begin{figure}[t]
\large
\begin{center}
\scalebox{0.9}{\includegraphics{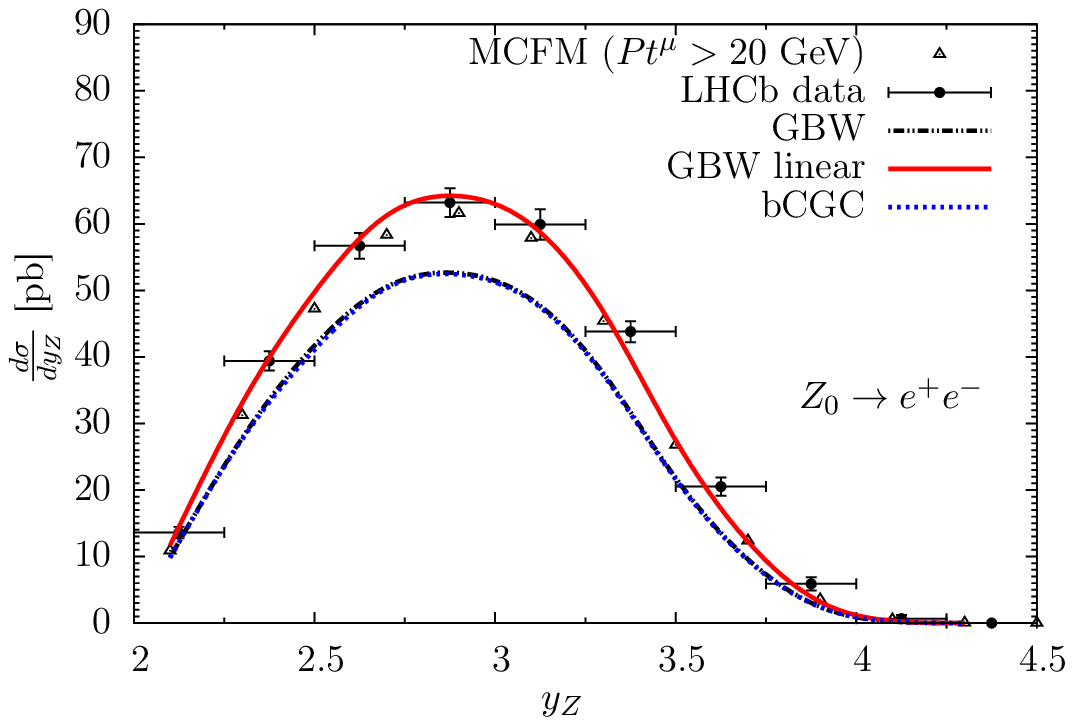}}
\caption{ (Color online) Rapidity distribution for the $Z$-boson production in the LHCb kinematical range.  Data from \cite{lhcb_data}.}
\label{fig:5}
\end{center}
\end{figure}

In Fig. \ref{fig:4} we present our predictions for the energy dependence of the $W$ and $Z$-boson cross sections obtained assuming that the gauge boson $G$ is produced in  the LHCb kinematical range ($ 2 \le \eta (G) \le 4.5$). For comparison we also present the MCFM predictions \cite{mcfm} which are obtained considering the NNLO corrections for the total cross sections. As already observed for the total cross sections, we obtain that the CGC and GBW predictions for the LHCb kinematical range are similar, being $\approx 20 \%$ smaller than the GBW Linear and MCFM one.  
The LHCb Collaboration has determined the $Z$ production cross section in three leptonic decay channels ($\mu^+ \mu^-, e^+ e^-, \tau^+ \tau^-$), with all events being selected by requiring the single muon or the single electron trigger. Lets initially compare our results with the experimental data for the rapidity distribution of the $Z$-boson and the corresponding MCFM predictions (See  Fig. \ref{fig:5}). We obtain that the GBW Linear prediction is very similar to the MCFM one and describe the data quite well. In contrast, the non-linear predictions underestimate the data at $y_Z \approx 3$. 
In order to compare our predictions with the recent LHCb data  for the cross sections \cite{lhcb_data}, it is necessary to apply the experimental cuts in the leptons in the final state. Following the procedure discussed in Refs. \cite{lhcb_data}, we obtain the results presented in  Table \ref{tab-2}, where we compare our predictions with  the LHCb data.
As expected from our previous analysis, the color dipole formalism describes the data quite well if we use the  GBW linear dipole - proton scattering amplitude. 

\normalsize

\normalsize

\begin{table}[h!]
\footnotesize 
\begin{center}
\begin{tabular}{|c@{\quad}||c@{\quad}|c@{\quad}|c@{\quad}|c@{\quad}|}
\hline

 & {\bf LHCb data}  & {\bf bCGC} & {\bf GBW} & {\bf GBW Linear}  \\ [0.5ex] \hline \hline

 $\sigma_Z Br(Z\rightarrow e^+ e^-)$      & 76.0 $\pm$ 0.8 $\pm$ 2.0 $\pm$2.6 & 59.35 & 58.91 & 73.61   \\ \hline

 $\sigma_Z Br(Z\rightarrow \mu^+\mu^-)$   & 76.7 $\pm$ 1.7 $\pm$ 3.3 $\pm$2.7 & 59.50 & 59.20 & 73.90  \\ \hline

 $\sigma_Z Br(Z\rightarrow \tau^+\tau^-)$ & 71.4 $\pm$ 3.5 $\pm$ 2.8 $\pm$2.5 & 59.50 & 59.20 & 73.90   \\ \hline

 $\sigma_{W^+} Br(W^+\rightarrow \mu^+\nu_{\mu})$     & 831 $\pm$ 9 $\pm$ 27 $\pm$ 29  & 659.07 & 667.59 & 836.78    \\ \hline

 $\sigma_{W^-} Br(W^-\rightarrow \mu^-\nu_{\mu})$     & 656 $\pm$ 8 $\pm$ 19 $\pm$23  & 599.19 & 593.90 & 710.23   \\ \hline
\end{tabular}
\end{center}
\caption{Comparison between the linear and non-linear predictions for the cross sections considering the LHCb experimental cuts. Data from  Ref.  \cite{lhcb_data}. Cross sections in pb. }
\label{tab-2}
\end{table}

A comment is in order here. As discussed before, the color dipole formalism used in our calculations disregard valence quark contributions as well as next-to-leading corrections. Both contribution can modify the normalization of the cross sections and rapidity distributions. A simplistic way to include these corrections is  multiplying the cross section by a $K$-factor, fixed by the data in order to obtain the correct normalization of the cross section. If it is made, we have checked that for a $K \approx 1.2$, independently of the energy, then the bCGC model is able to describe the data for the total cross section as well as the LHCb data for the rapidity distributions. However, it is important to emphasize that the inclusion of a common  value for the $K$-factor, independent of the gauge boson produced, do not improve the description of the ratio between the $W$ and $Z$ cross section presented in Fig. \ref{fig:4}. Therefore, this subject deserve more detailed studies. 


\section{Conclusions}
\label{conclusoes}

The study of the production of the massive gauge bosons $W^{\pm}$ and $Z^0$ in proton - proton collisions  provide an important test of the  perturbative QCD. The description of this process is usually made using the collinear factorization and taking into account the  perturbative contributions  up to next-to-next-to-leading order, with the corresponding predictions describing the data quite well. However,  the current and future experimental data for the gauge boson production also provide an important test of the QCD dynamics at high energies (small-$x$), where  gluon saturation effects can be present. Such effects are  naturally described in the color dipole formalism. This formalism also ressums leading logarithms in the energy and takes into account higher twist contributions for the cross sections, which are disregarded by the collinear factorization approach. Thus, no matter if the gluon saturation effects are or not important for the gauge boson production, it is worth to obtain a description of the $W^{\pm}$ and $Z^0$ cross sections in this framework.  The first step was performed in Ref. \cite{pkp} generalizing previous analysis of the Drell-Yan process in the color dipole formalism \cite{k95,kts99,brodsky,krt01}. Our goal in this paper was to improve that analysis, performing a systematic study of the inclusive production considering different models for the dipole - proton cross section and a detailed  comparison of its predictions with the current experimental. Through the comparison between the full and linear predictions of the GBW model, we have  obtained that the gluon saturation effects contribute for the gauge boson production at LHC energies by approximately 20\%. However, our results indicate that  current experimental data for the gauge boson production can only be  described in the color dipole formalism if the non-linear effects are disregarded.
In particular, the  bGCC model, which successfully describes 
the recent high precision combined HERA data, fails to describe the recent ATLAS, CMS and LHCb data.
In principle, it can be associated to the fact that the bCGC model disregard the DGLAP evolution in the description of the dipole - target cross section at small values of the dipole pair separation, which dominates the gauge boson production. Another possible interpretation is that valence quark contributions and/or higher orders corrections to the formalism should be taken into account. Both possibilities deserve more detailed studies, which we plan to perform in the future. Finally, as the saturation scale increases with the atomic number it will be interesting to generalize our results for proton - nucleus collisions, estimating  the gauge boson production in this process at LHC energies and make a comparison between our predictions with those which come from the collinear factorization with nuclear effects included in the nuclear parton distributions \cite{salgado}.

\section*{Acknowledgements}
 We would like to thank Roman Pasechnik for very fruitful discussions about this and related subjects. V. P. G. is grateful to the Theoretical High Energy Physics Group at Lund University for support and hospitality during final stages of this work. E. A. F. Basso is supported by CAPES (Brazil) under contract 2362/13-9. This work has been supported by CNPq, CAPES and FAPERGS, Brazil.
 
\bibliographystyle{unsrt}
 
\end{document}